%% file: template.tex
\documentclass[onecolumn, journal]{IEEEtran}
\input{package}

\ifCLASSINFOpdf
\else
\fi

\begin{document}
%
\title{Design and Comparison of Reward Functions in Reinforcement Learning for Energy Management of Sensor Nodes}
%
%
%





\author{Yohann~Rioual,
        Yannick~Le Moullec,
        Johann~Laurent,
        Muhidul~Islam Khan,
        and Jean-Philippe~Diguet
\thanks{Y. Rioual and J. Laurent are with Lab-STICC, University Bretagne Sud, F-56100 Lorient, France, e-mail: \{firstname.lastname\}@univ-ubs.fr}
\thanks{Y. Le Moullec and M. Islam Khan are with Thomas Johann Seebeck Department of Electronics, Tallinn University of Technology, Ehitajate tee 5, 19086 Tallinn, Estonia; yannick.lemoullec@taltech.ee, mdkhan@taltech.ee}
\thanks{J-Ph. Diguet is with IRL CNRS CROSSING, Adelaide, SA, Australia}%
\thanks{This paper is an extended version of our paper published in 2018 16th Biennial Baltic Electronics Conference (BEC).}%
}

\maketitle

\begin{abstract}
\input{Paper/abstract}
\end{abstract}

\begin{IEEEkeywords}
reinforcement learning, IoT, WBANs, reward function, Q-learning, energy management
\end{IEEEkeywords}

%
\IEEEpeerreviewmaketitle

\input{Paper/intro}

\input{Paper/related}

\input{Paper/RL}

\input{Paper/WBAN}

\input{Paper/design}

\input{Paper/conclusion}

\ifCLASSOPTIONcaptionsoff
  \newpage
\fi



%
\bibliographystyle{IEEEtran}
\bibliography{biblio}

%







\end{document}

%% file: package.tex
\usepackage{amsmath, amsthm, amssymb, amsfonts}
\usepackage{algorithmic}
\usepackage{algorithm}
\usepackage{graphicx}
\usepackage{array,multirow}
\usepackage{siunitx}

\usepackage{cleveref}

%% file: Paper/abstract.tex
Interest in remote monitoring has grown thanks to recent advancements in Internet-of-Things (IoT) paradigms. New applications have emerged, using small devices called sensor nodes capable of collecting data from the environment and processing it. However, more and more data are processed and transmitted with longer operational periods. At the same, the battery technologies have not improved fast enough to cope with these increasing needs. This makes the energy consumption issue increasingly challenging and thus, miniaturized energy harvesting devices have emerged to complement traditional energy sources. Nevertheless, the harvested energy fluctuates significantly during the node operation, increasing uncertainty in actually available energy resources. Recently, approaches in energy management have been developed, in particular using reinforcement learning approaches. However, in reinforcement learning, the algorithm's performance relies greatly on the reward function. In this paper, we present two contributions. First, we explore five different reward functions (\ref{eq:1}--\ref{eq:5}) to identify the most suitable variables to use in such functions to obtain the desired behaviour. Experiments were conducted using the Q-learning algorithm to adjust the energy consumption depending on the energy harvested. Results with the five reward functions illustrate how the choice thereof impacts the energy consumption of the node. Secondly, we propose two additional reward functions (\ref{eq:rewardConf} and \ref{eq:rewardConf2}) able to find the compromise between energy consumption and a node performance using a non-fixed balancing parameter. Our simulation results show that the proposed reward functions (\ref{eq:rewardConf} and \ref{eq:rewardConf2}) adjust the node's performance depending on the battery level and reduce the learning time.

%% file: Paper/intro.tex
\section{Introduction}
\IEEEPARstart{T}{he} rapidly growing interest for physiological sensors and progress in low-power integrated circuits and wireless communication have enabled a new generation of wireless sensor networks (WSN). WSNs consist of a number of communicating nodes, they can be fitted on the human body to monitor physiological parameters of the wearer or environmental variables. Advances in microelectronics led to significant miniaturization of the sensors; however, battery technologies have not improved at the same rate \cite{yuan2011lithium}. Consequently, the storage of energy is a bottleneck for the deployment of such sensors.

To minimize the battery's size, an increasingly popular approach is to harvest energy from the environment \cite{zhou2014harvesting}. Miniaturised energy harvesting technologies cannot harvest a lot of energy (see Table \ref{HT}), but they can be used as complements to the battery. However, such energy sources vary greatly over time and bring uncertainties in the system energy resources. Reinforcement Learning (RL) algorithms have acquired a certain popularity in recent years for energy management (\cite{shresthamali2017adaptive},~\cite{ahmed2017qos}). Indeed, they can handle such uncertainty in energy sources and appear to be a valid solution for energy management. They adapt the node's behaviour by promoting good decisions by means of a reward function \cite{kaelbling1996reinforcement}. 

\begin{table}[ht]
\center
\caption{\label{HT} Power density of energy harvesting technologies \cite{kim2014ambient}}
\resizebox{.6\linewidth}{!}{%
\begin{tabular}{l|c}
Harvesting technologies & Power density\\\hline\hline
Solar cell (outdoors at noon) & $15~\si{\milli\watt/\centi\meter\squared}$ \\
Wind flow (at $5~\si{\meter/\second}$) & $16.2~\si{\micro\watt/\centi\meter\cubed}$\\
Vibration (Piezoelectric -- shoe insert) & $330~\si{\micro\watt/\centi\meter\cubed}$\\
Vibration (electromagnetic conversion at $52~\si{\hertz}$) & $306~\si{\micro\watt/\centi\meter\cubed}$\\
Thermo-electric ($5~\si{\celsius}$ gradient) & $40~\si{\micro\watt/\centi\meter\cubed}$\\
Acoustic noise ($100~\si{\dB}$) & $960~\si{\nano\watt/\centi\meter\cubed}$
\end{tabular}
}
\end{table}

The choice of an appropriate reward function is challenging. Since this function determines the behaviour of the system, choosing it is an essential task for the system designer. Still, the literature on this topic rarely discusses the choice of the reward function.

Thus, in this paper, we explore the influence of different parameters on the reward function performance with a popular RL algorithm, i.e. Q-learning. This paper complements the work presented in \cite{rioual2018reward}, we present two different contributions:
\begin{itemize}
\item[$\bullet$] a comparison of five different reward functions (\ref{eq:1}--\ref{eq:5}) to identify the most suitable variables to design a function able to adjust correctly the energy consumption depending of the harvested energy. This version of the paper provides some additional details as compared to \cite{rioual2018reward}.  
\item[$\bullet$] using the previous results, we expand \cite{rioual2018reward} with the detailed design of two additional reward functions (\ref{eq:rewardConf} and \ref{eq:rewardConf2}) which find a compromise between the energy consumption and node's performance depending on the battery level with a reduced learning time.
\end{itemize}

The remainder of this paper is structured as follows. \Cref{related_work} discusses the related work. \Cref{reinforcement_learning} introduces the RL mechanism and presents the Q-learning algorithm used in this study. \Cref{experiments} presents a body sensor node as use case and experimental results wherein five reward functions are evaluated and compared. Then, \Cref{design} presents a piecewise reward function which adjusts the energy consumption of a sensor node according to the battery state of charge. \Cref{continuous} presents a generalization of the previous function. Lastly, \Cref{conclusion} summarizes and concludes this paper.

%% file: Paper/related.tex
\section{\label{related_work}Related work}

Works on energy management in sensor networks such as WBANs typically focus on the consumption of the radio, often disregarding the power consumption of the other parts of the system. In cases where energy harvesting technologies are employed, related works have proposed adaptive protocols that deal with the challenge of providing the required quality of service under the uncertainty of the energy input provided by a human activity. \cite{ibarra2016qos} presents a method for adapting the transmission of data according to the energy harvested. The measured data is stored in a queue and it is transmitted if there is enough available energy. The old data that lost their validity, determined by the medical application requirements, are deleted from the queue to avoid having to transmit it and preventing a buffer overflow. If the queue is full then the oldest data is deleted. The authors also propose an algorithm that determines, during the communication, how many packets will be transmitted according to the available energy and the state of the queue.

To extend a sensor node's lifespan in WBANs, there also exist different energy management methods. This includes adapting the node's energy consumption to the activity of the wearer. To this end, \cite{casamassima2014context} proposes a classifier to identify the activity of a wearer using data collected from an accelerometer and adapts the operating policy accordingly. In the literature, there is a variety of classifiers, so the authors compare five different approaches (Nearest Neighbour, Support Vector Machines, Naïve Bayes, Linear Discriminant and Decision Tree). Their comparison criterion is to minimize the trade-off between computational cost, power consumption and recognition performance. The decision tree method corresponds the most to the application requirements and detects the activity of the wearer with an accuracy of $98\%$. The output of the classifier is used to change the node's energy policy according the activity and thus to adapt the energy consumption. The authors tested it in a use case scenario with continuous monitoring and feedback exercises. Their results show a five-time increase in battery lifetime. 

Since RL is an approach to take decisions under uncertainty, it is suitable for energy management in systems where energy harvesting technologies are used. For example, the authors of \cite{Kazemi2011} have developed a power control approach in WBANs based on RL. This approach mitigates the interference due to wireless communication in the network, which provides a substantial saving in energy consumption per transmitted bit. However, this energy management approach only focuses on the wireless communication and not on the other parts of the WBAN node. Communications represent an important part of the energy consumption in WBANs, nevertheless recent trends increase the computational load, such as data-preprocessing or edge computing on the node to reduce the energy consumed by communication, which increases the energy consumed by sampling and processing.

The energy management of a sensor node is a complex problem. On the one hand, the aim is to reduce the energy consumption, on the other hand the performance must be maximized or satisfied depending on the application constraints. A solution to this emerges with Multi-Objective Reinforcement Learning (MORL)~\cite{liu2015multiobjective}. The aim of MORL is to solve problems where there are multiple conflicting objectives. The naive approach consists of a look-up table for each objective, the choice being made on the action that maximizes the sum of the different Q-values. 

There also exist different non-naive algorithms such as weighted sum \cite{karlsson1997learning} and W-learning \cite{humphrys1996action} for single policy approaches, but they cannot express exactly the preferences of the designer. If the preferences must be expressed more exactly, there exist Analytic Hierarchy Process (AHP) \cite{zhao2010multi}, ranking \cite{vamplew2011empirical} and geometric \cite{mannor2004geometric} approaches but they need prior knowledge of the problem domain. For a multi-policy approach, there exist the convex hull algorithm \cite{barrett2008learning}. However, some researchers argue that explicit multi-objective modelling is not necessary, and that a scalar reward function is adequate \cite{sutton1998reinforcement}. For instance Sutton’s reward hypothesis, states “that all of what we mean by goals and purposes can be well thought of as maximization of the expected value of the cumulative sum of a received scalar signal (reward)”. The implication is that a multi-objective Markov Decision Process can always be converted into a single-objective one. We will therefore focus on designing reward functions that take into account different parameters in a single objective, i.e., reduction in energy consumption.

There are few papers about WBANs that deal with harvesting energy and RL for the energy management of the entire node (\cite{ahmed2017qos}, \cite{chen2019reinforcement}, \cite{jagannath2019machine}). The process of designing a reward function is not explained in most papers; for instance \cite{Kazemi2011} shows good results with one reward function without explaining if they tried different reward functions. The rest of the papers apply different RL approaches to various problem and do not discuss the construction of the reward function they used. The definition of a reward function is important and depends on the application. The following section introduces the reinforcement learning and one algorithm in particular, the Q-learning.

%% file: Paper/RL.tex
\section{\label{reinforcement_learning}Reinforcement Learning}

In this section, we give an overview of RL. Then we present the selected Markov Decision Process (MDP) used to deal with the energy management process. And to conclude this section, we introduce the selected RL algorithm, i.e. Q-learning.

\subsection{Overview of Reinforcement Learning}
\begin{figure}[h]
\centering
\includegraphics[width=.6\linewidth]{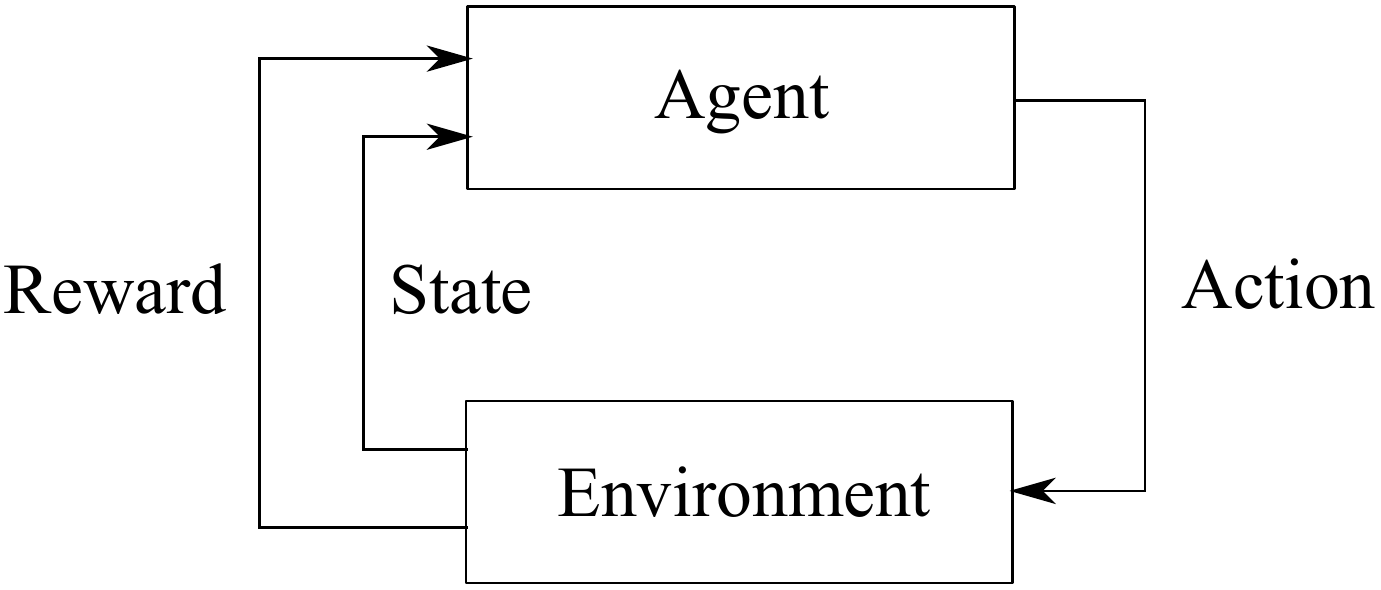}
\caption{\label{RL}Interaction between an agent and its environment}
\end{figure}

RL is a formal framework that models sequential decision problems \cite{sutton1998reinforcement}, in which an agent learns to make better decisions by interacting with the environment (\Cref{RL}). When the agent performs an action, the state changes and the agent receives a feedback called a reward, which indicates the quality of the transition. The agent's goal is to maximize its total reward over the long-term.

There is a trade-off between exploration and exploitation in RL. During the exploration, the agent chooses an action randomly to find out the utility of that action, whereas during the exploitation, the action is chosen based on the previously learned utility of the actions. The exploration probability is given according to \Cref{eq:heuristic} \cite{khan2014energy}:
\begin{equation} \label{eq:heuristic}
\epsilon = min(\epsilon_{max}, \epsilon_{min} + k \times (S_{max} - S)/S_{max})
\end{equation}
where $\epsilon_{max}$ and $\epsilon_{min}$ denote upper and lower limits of the exploration factor, respectively, $S_{max}$ represents the size of the state space, and $S$ represents the current number of states already visited. At each time step, the system computes $\epsilon$ and generates a random number in the interval $[0, 1]$. If the selected random number is at most $\epsilon$, the system chooses a uniformly random action (exploration), otherwise it chooses the best one using $Q$-values (exploitation).

\subsection{Markov Decision Process for Energy Management}
The energy management problem can be modelled as a Markov Decision Process (MDP). An MDP provides a theoretical framework for modelling decision making in situations where outcomes are partly random and partly under the control of a decision maker. An MDP is formally defined as a n-tuple $\left< \mathcal{S},\mathcal{A},\mathcal{T},\mathcal{R} \right>$ where $\mathcal{S}$ is a state space, $\mathcal{A}$ a set of possible actions, $\mathcal{T} : \mathcal{S} \times \mathcal{A} \times \mathcal{S} \rightarrow [0,1]$ are the transitions' probabilities between states $\left(\mathcal{T}(s,a,s') = p(s^\prime|a,s)\right. $ is the probability to reach state $\left.s' \text{ starting from } s \text{ after taking an action } a \right)$, and $\mathcal{R} : \mathcal{S} \times \mathcal{A} \rightarrow \mathbb{R}$ is a reward signal.

The transitions' probabilities between the states $T$ are unknown, so we use a model-free algorithm. Model-free algorithms work even when we do not have a precise model of the environment. These algorithms primarily rely on learning algorithms such as Q-learning, which is described in the next section.

\subsection{Q-learning algorithm}
In this section, we present the Q-Learning algorithm \cite{watkins1992q}. The Q-learning algorithm is widely used thanks to its ease of implementation yet effective performance, and its convergence is proven. We use Algorithm~1 for the energy management of a sensor node in combination with the MDP presented above.

\begin{algorithm}[ht]
    \caption{\label{alg:Q} Q-learning algorithm}
    \begin{algorithmic}
    \STATE Initialise $Q(s,a)$ arbitrarily
    \STATE The agent observes the initial state $s_{0}$
	\FOR{each decision epochs}
	\STATE Choose an action $a$ from state $s$ using policy derived from $Q$
	\STATE Take action $a$, observe the new state $s'$ and the associated reward $r$
	\STATE Update of the related $Q$-value:\newline $Q(s,a)\leftarrow Q(s ,a) + \alpha\left(r+\gamma \max\limits_{a'}Q(s', a')-Q(s, a)\right)$
	\STATE $s \leftarrow s'$
	\ENDFOR
    \end{algorithmic}
\end{algorithm}

\subsubsection*{Learning rate $\alpha$} The learning rate $\alpha$ determines how fast the new information will replace the old one. A factor of 0 would not teach the agent in question anything, whereas a factor of 1 would only teach the agent the latest information. In our work, we slowly decrease the learning rate $\alpha$ in such a way that it reflects the degree to which a state-action pair has been chosen in the past (\Cref{eq:alpha}).
\begin{equation} \label{eq:alpha}
\alpha = \frac{\zeta}{visited(s,a)}
\end{equation}
where $\zeta$ is a positive constant and $visited(s,a)$ represents the visited state-action pairs so far \cite{bianchi2004advances}.

\subsubsection*{Discount factor $\gamma$} The discount factor $\gamma$ determines the importance of future rewards. A discount factor of 0 would make the agent myopic by considering only current rewards, while a factor close to 1 would also involve more distant rewards. If the discount factor is close or equal to 1, the value of Q may never converge.

Using the MDP, we aim to identify the best variables to use in the reward function to adapt the energy consumption according to the energy we can harvest. In the following section, we present some results and identify those variables.

%% file: Paper/WBAN.tex
\section{\label{experiments}Energy management of a body sensor node}

In this section, we test five reward functions (R1 to R5) to identify the most valuable one for the energy management of a body sensor node. First, we present the use case and the decision process, then we present results of the node's behaviour. 

\subsection{Use case}
\begin{figure}
    \centering
    \includegraphics[width=.4\linewidth]{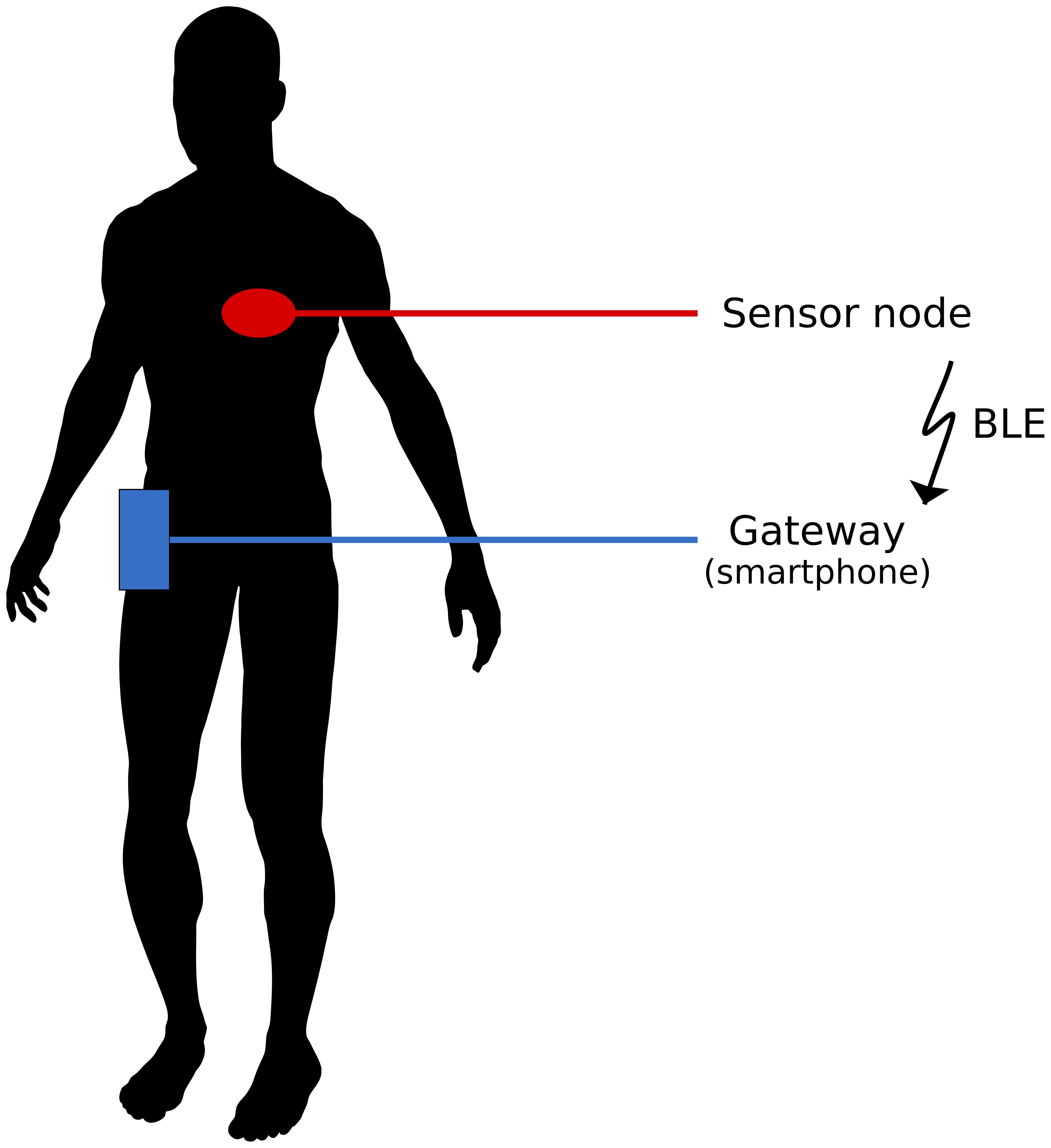}
    \caption{Sensors node fitted on a chest to monitor the heart beat}
    \label{fig:use_case}
\end{figure}

In this work, the objective is to manage the energy consumption of a sensor node fitted on a human chest to monitor the cardiac activity for non-medical application (\Cref{fig:use_case}). The heart rate is measured for 10 seconds. Then, data are sent immediately to a smartphone to be processed. The smartphone is used as a gateway and communicates with the node using a Bluetooth Low Energy (BLE) transceiver. The node does not continuously monitor the cardiac activity, i.e. after each measurement it enters a sleep mode to minimize energy consumption. The period between each measurement is variable and lasts from 10 to 60 minutes.

The sensor node is equipped with an optical heart rate detection sensor, a low-power micro-controller unit (MCU), a BLE transceiver and a battery with a capacity of $100~\si{\milli\ampere\!\hour}$. The energy consumption of each component is summarized in \Cref{table:conso}. The energy consumption of the MCU depends on the processor clock frequency. Here, the maximum frequency is $32~\si{\mega\hertz}$, corresponding to a maximum current consumption of $7.2~\si{\milli\ampere}$. As can see in \Cref{table:conso}, the communication unit (BLE transmitter) in active mode consumes more than the two other components combined.

\begin{table}[h]
\center
\caption{\label{table:conso}Node components and respective current consumption}
\resizebox{.5\linewidth}{!}{
\begin{tabular}{c|c|c}
Component & Active mode & Sleep mode\tabularnewline
\hline \hline
Heart rate sensor & $1.6~\si{\milli\ampere}$ & $0.12~\si{\milli\ampere}$\tabularnewline
Micro-controller & $225~\si{\micro\ampere}/\si{\mega\hertz}$& $0.93~\si{\micro\ampere}$\tabularnewline
BLE transmitter & $10.5~\si{\milli\ampere} $ & $0~\si{\micro\ampere}$  (turned off)\tabularnewline
\end{tabular}}
\end{table}

To increase the amount of energy available, a kinetic motion energy harvester is added to the battery. This energy harvester is presented in \cite{gorlatova2014movers}. The harvested energy is low to power entirely a node, but it still can extend the node's lifespan; Table \ref{harvester} shows how much power can be harvested according to the activity of the wearer. These data are extracted from \cite{gorlatova2014movers}.

\begin{table}[h]
\center
\caption{\label{harvester}Kinetic motion's harvested power for three different activities}
\resizebox{.25\linewidth}{!}{
\begin{tabular}{c|c}
Activity & Power harvested\tabularnewline
\hline\hline
relaxing & $\num{2.4}~\si{\micro\watt}$\tabularnewline
walk & $\num{180.3}~\si{\micro\watt}$\tabularnewline
run & $\num{678.3}~\si{\micro\watt}$
\end{tabular}}
\end{table}

We use the dominant frequency of motion, $F_m$, to identify which activity is performed by the wearer. We obtain $F_m$ by determining the maximum spectral component of the Fourier Transform of the acceleration $a(t)$. Since the harvested energy is uncertain, we use an RL approach to manage the node's consumption by adjusting its sleep duration and its processor's clock frequency.

\subsection{Decision Process}
For this use case, we define a set of actions with different processor frequencies ($F_p$) and periods between each measurement ($P_s$) (Table~\ref{table:actions}). For instance, Action $1$ has a processor frequency of $32~\si{\mega\hertz}$ and a measurement every minute, whereas Action 3 has a processor frequency of $4~\si{\mega\hertz}$ and a measurement every 5 minutes. Thus, Action 1 consumes more current (and thus more energy) than Action 3. All actions have different energy consumption levels since they depend on the processor's frequency in active mode and its consumption in sleep mode (see the third row in Table~\ref{table:actions}).

\begin{table}[h]
\center
\caption{\label{table:actions}Set of actions with different processor frequencies ($F_p$) and periods between each measurement ($P_s$), and the associated average current consumption}
\resizebox{.5\linewidth}{!}{
\begin{tabular}{c|c|c|c}
Action & $F_p$ & $P_s$ & Average current consumption \\
\hline \hline
1 & $32$ \si{\mega\hertz} & $1$ \si{\minute} & $0.6278$ \si{\milli\ampere}\\
2 & $4$ \si{\mega\hertz} & $1$ \si{\minute} & $0.4873$ \si{\milli\ampere}\\
3 & $4$ \si{\mega\hertz} & $5$ \si{\minute} & $0.2292$ \si{\milli\ampere}\\
4 & $4$ \si{\mega\hertz} & $20$ \si{\minute} & $0.2044$ \si{\milli\ampere}\\
5 & $1$ \si{\mega\hertz} & $60$ \si{\minute} & $0.1926$ \si{\milli\ampere}
\end{tabular}}
\end{table}

Our state space is divided into three different states corresponding to the activity of the wearer (\Cref{harvester}). We use the dominant frequency motion $F_m$ which is correlated with the energy we harvest to consider our state. A high value of $F_m$ corresponds to more energy being harvested and a low value of $F_m$ corresponds to less energy being harvested. The state is determined with the value of $F_m$ and corresponds to an activity. The activity can be considered high (i.e. running) if $F_m > 2~\si{\hertz}$ , moderate (i.e. walking) if $2~\si{\hertz} \geq F_m > 1~\si{\hertz}$ or low (i.e. relaxing) if $F_m \leq 1~\si{\hertz}$.

\subsection{Simulation results}
First of all, it should be noted that the harvesting capabilities of the kinetic motion harvester are not sufficient to fully recharge the sensor node's battery. Thus, we seek and expect to reduce the node's consumption when the harvested energy is low. We test five different reward functions to identify which parameters have an appropriate influence on our system's behaviour. To avoid divergence in the Q-values, the values of the different reward function are limited to [-1, 1].

There are different constraints when designing the system and most of them are conflicting; for instance keeping the sleep period as short as possible while also reducing energy consumption. The main purpose of the RL algorithm is to find the equilibrium point to respect these constraints. To this end, the first and second reward functions use a parameter $\beta$ to balance the equilibrium point according to what is considered most important between performance and battery level \cite{khan2014energy}. 

The first reward function  (\ref{eq:1})  seeks to balance the conflicting objectives between the sleep duration $P_s$ and the energy consumption of the sensor node. $B_r(t)$ is the residual energy in the battery's node at time $t$.
\begin{equation} \tag{R1} \label{eq:1}
R = \beta*\frac{\min(P_s)}{P_s} + (1 - \beta)*(B_r(t) - B_r(t-1))
\end{equation}

The second reward function  (\ref{eq:2})  is similar to the first one but instead of using the energy consumption, it only uses the residual energy of the battery's node at time $t$.
\begin{equation} \tag{R2} \label{eq:2}
R = \beta*\frac{\min(P_s)}{P_s} + (1 - \beta)*\frac{B_r(t)}{B_{max}}
\end{equation}

The third reward function (\ref{eq:3})  does not consider the sleep duration $P_s$ but only the energy consumption. The objective is to find the less consuming operating mode without taking care of the performance.
\begin{equation} \tag{R3} \label{eq:3}
R = B_r(t) - B_r(t-1)
\end{equation}

Finding the right setting for $\beta$ is not trivial, that is why the fourth reward function (\ref{eq:4}) uses the product of the sleep duration $P_s$ and the residual energy $B_r(t)$. Indeed, the result is maximal when both values are maximum.
\begin{equation} \tag{R4} \label{eq:4}
R = \frac{\min(P_s)}{P_s} \times B_r(t)
\end{equation}

We want to adapt the energy consumption to the activity of the wearer. So, in the fifth reward function (\ref{eq:5}), we use the dominant motion frequency $F_m$ which determines the activity. The aim is to minimize the difference between the normalized $F_m$ and the energy consumption; a cosine function restricts the reward to $[-1,1]$. 
The reward is maximized when the difference is close to $0$. Moreover, this reward function eliminates the $\beta$ parameter that is not trivial to adjust. $\mathcal{N}$ is the normalization function.
\begin{equation} \tag{R5} \label{eq:5}
R = \cos\left(\frac{\mathcal{N}(F_m) - (B_r(t)-B_r(t-1))}{2}\right)
\end{equation}

\begin{figure*}[htbp]
\center
\includegraphics[width=0.75\linewidth]{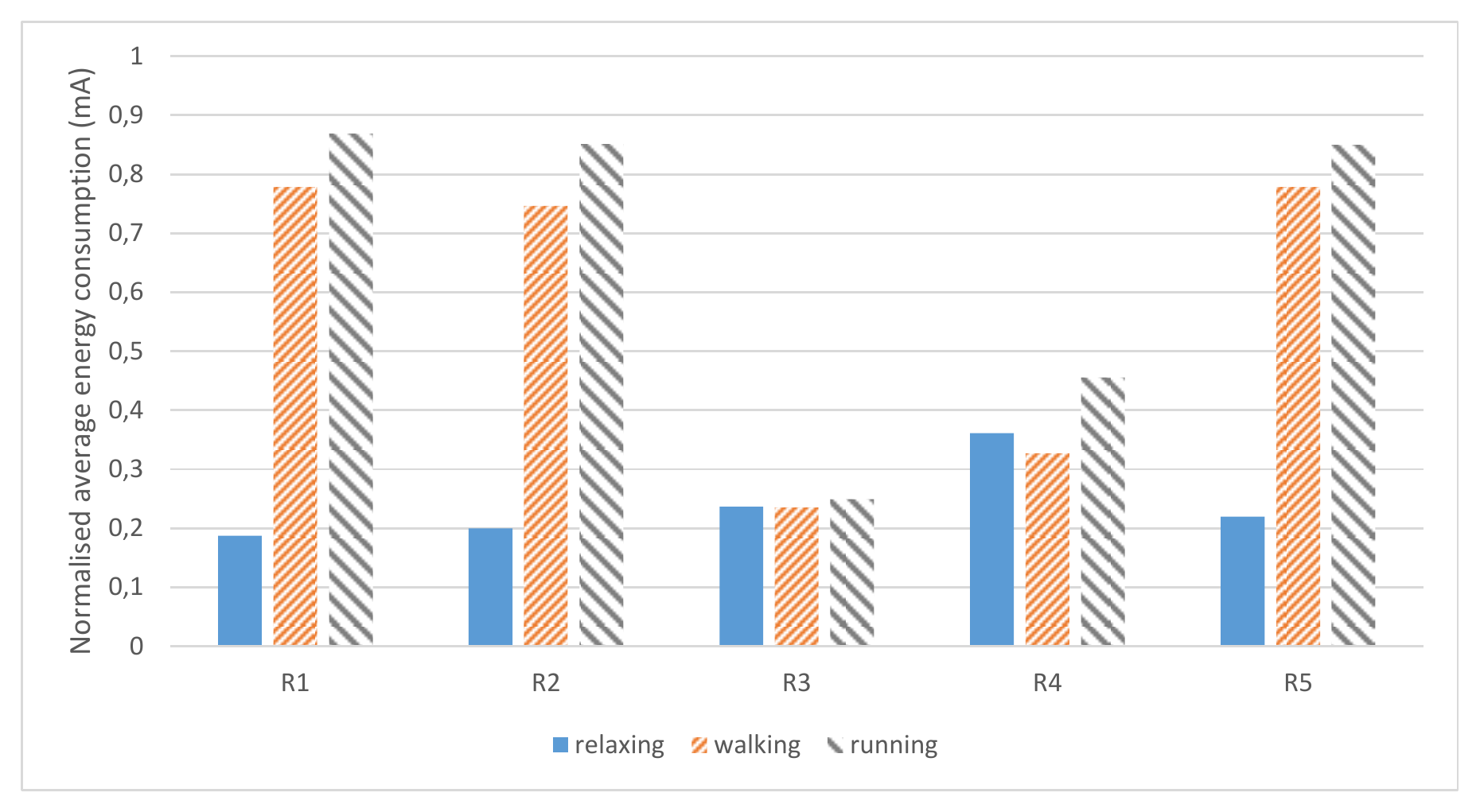}
\caption{\label{fig:figure} Normalised average energy consumption of the node according to 1) the activity of the wearer and 2) the reward function used within the Q-learning algorithm. \ref{eq:1}, \ref{eq:2} and \ref{eq:5} behave as expected since they allow the node to consume more when more energy is harvested.}
\end{figure*}

We simulate a week-long node deployment to observe the evolution of the battery's charge level. The activity changes every 30 minutes, and our agent chooses an action (\Cref{table:actions}) every 20 minutes. \Cref{fig:figure} shows the average energy consumption of the node according to the activity identified with the dominant frequency of motion, $F_m$. The parameter $\beta$ is fixed at $0.3$ since our primary goal is to adapt the node's consumption, i.e. we give more importance to the energy factor. The results show that the choice of the reward function has a significant impact on the average current consumption. While some reward functions yield the expected behaviour, others adapt poorly to the wearer's activity and others do not yield the correct behaviour at all, as discussed in what follows.

The expected behaviour of the node is to adjust its energy's consumption depending on the harvested energy. Thus, during a physical effort, when the harvested energy is high, the node realizes more measurements. A second objective is that the node needs to survive at least a week to reduce the number of human intervention on the node. This second objective is achieved for all the reward functions. Based on \Cref{fig:figure}, our observations are as follows: 

$\blacktriangleright$ Reward function \ref{eq:1} computes the reward using the sleep time and the energy consumption of the node. This function produces a maximal value when the sleep time and the energy consumption are low. It successfully adapts the energy consumption according to the activity, increasing the node's consumption when the harvested energy increases. A drawback is how to select the most appropriate value for the $\beta$ parameter.

$\blacktriangleright$ Reward function \ref{eq:2} computes the reward with the sleep time $P_s$ and the battery's residual energy. In the same way as the reward function \ref{eq:1}, it successfully adapts the energy consumption according to the activity, achieving lowest energy consumption than the reward function \ref{eq:1} in 2 activities (walking and running). Furthermore both rewards share the same drawback, the choice of the value of the $\beta$ parameter.

$\blacktriangleright$ Reward function \ref{eq:3} computes the reward only using the node's consumption. It fails to adapt the node's behaviour according to the harvested energy. It does not make any difference between the activities; however, it succeeds to minimize the energy consumption. At the end of the simulation, the battery charge is still above $75\%$ .

$\blacktriangleright$ Reward function \ref{eq:4} computes the reward with the product of the $P_s$ and the battery's residual energy. This reward does not have a parameter to tune, and it is easy to compute. Unfortunately, it fails to adjust the node's consumption according to the harvested energy. The reward function increases the node's consumption when the wearer is relaxing (i.e. the energy harvested is low), decreases it when the wearer is walking and then increases the node's consumption when the wearer is running (i.e. the energy harvested is maximal). This is obviously not desired and the reward function (\ref{eq:4}) is more influenced by the sleep time $P_s$ than by the consumption of the sensor node.

$\blacktriangleright$ Reward function \ref{eq:5} computes the reward with the normalized value of the dominant frequency of motion $F_m$ and the node's consumption. The reward is maximal when the difference between the energy consumption and the normalized dominant frequency of motion is close to $0$. The function \ref{eq:5} fulfills the different objectives we had set at the beginning. There is not a parameter to tune, and so, this reward function can be used easily in this application. Nevertheless, the absence of this parameter makes this function less relevant for other applications with different requirements.

\begin{table}[h]
    \center
    \caption{\label{table:eval}Evaluation of the different reward functions}
    \resizebox{.7\linewidth}{!}{%
    \begin{tabular}{c|ccc}
        Reward function & Configurable & Energy consumption & Compliance with the objectives \\ \hline \hline
        \ref{eq:1} & $\bigstar\bigstar\bigstar$ & $\bigstar\bigstar$ & $\bigstar\bigstar\bigstar$ \\
        \ref{eq:2} & $\bigstar\bigstar\bigstar$ & $\bigstar\bigstar$ & $\bigstar\bigstar\bigstar$ \\
        \ref{eq:3} & -- & $\bigstar\bigstar\bigstar$ & -- \\
        \ref{eq:4} & -- & $\bigstar$ & -- \\
        \ref{eq:5} & -- & $\bigstar\bigstar$  & $\bigstar\bigstar\bigstar$ 
    \end{tabular}}
\end{table}


The overall evaluation of the five reward functions is summarized in \Cref{table:eval}. The reward functions \ref{eq:1} and \ref{eq:2} allows regulating the importance given to the energy consumption according to the application requirements by increasing or decreasing the value of $\beta$, whereas the reward functions (\ref{eq:3}) and (\ref{eq:4}) are not relevant to adapt correctly the energy in a sensor node. The correct behaviour of the node can be obtained by using a $\beta$ parameter to balance energy consumption and performance (\ref{eq:1}, \ref{eq:2}). However, it is necessary to adjust this parameter. Using the dominant motion frequency in \ref{eq:5} removes this parameter and still achieves the right behaviour. However, this reward function is less modular. It allows adapting the energy consumption according to the activity but does not take the sleep duration into account.

IoT sensors are also used in other applications such as monitoring environmental conditions. For this kind of application, we could prefer to reduce performance when the battery is low and then increase it after it has recharged. In the following section, we present different designs for reward functions to balance the performance according to the residual energy in the battery. For this, we use a monitoring application based on a marine buoy.

%% file: Paper/design.tex
\section{\label{design}Design of a Piecewise Reward Function}
As seen in the previous section, the use of a balancing parameter allows a designer to design a reward function that complies with the objectives. Moreover, the use of a balancing parameter makes the reward function configurable, either to maximize a performance parameter or to preserve the battery's energy. 

The balancing parameter is selected using experience of the designer and it is fixed. We propose a reward function in which the configuration changes depending on the battery level (Equation \ref{eq:rewardConf}). 

\begin{equation} \tag{R6}
\label{eq:rewardConf}
R  = \left\{   
    \begin{array}{cl}
        F_s\times\rho_1 + B\times(1-\rho_1)&B\geq 75\%\\
        F_s\times\rho_2 + B\times(1-\rho_2)&75\%>B\geq 50\%\\
        F_s\times\rho_3 + B\times(1-\rho_3)&50\%>B\geq 25\%\\
        F_s\times\rho_4 + B\times(1-\rho_4)& \text{otherwise}
   \end{array}
   \right .
\end{equation}
where $1\geq\rho_1>\rho_2>\rho_3>\rho_4\geq 0$ are balancing parameters, $F_s$ the sampling frequency and $B$ is the charge of the battery.

When the battery is fully charged, it is useless to preserve it and it is possible to maximize the performance. Whereas, when the battery is discharged, it becomes really important to restrict the energy consumption in order to recharge the battery and extend the node's lifetime. A difficulty with this reward function is its adjustment. In Equation \ref{eq:rewardConf}, the battery level is divided into four equal-sized parts; however, this may vary according to the application or the node. Moreover, the parameters $\rho_{1,2,3,4}$ must be selected and a poor choice will make the reward function less effective or even not effective at all.  

To evaluate the proposed reward function, a simulation is conducted for the deployment of a marine buoy near Lorient, France, for 21 days. The buoy is equipped with two sensors (Table \ref{tab:component}), a 3D anemometer and an atmospheric sensor, to monitor environmental conditions. The sensors have different energetic behaviours and the buoy should be deployed for as long as possible. To complete the battery, the buoy is equipped with two small solar panels. To avoid collision with ships, a beacon light flashes for $500~\si{\milli\second}$ every four seconds when the brightness is low.

\begin{table}[ht]
\center
\caption{\label{tab:component} Buoy components}
\resizebox{.35\linewidth}{!}{
\begin{tabular}{l|c}
\textbf{Components} & \textbf{Device}\\ \hline\hline
3D Anemometer & WindMaster HS\\	 \hline
Atmospheric sensor & YOUNG61302L\\ \hline
Processor & Cortex-M4 MCU\\ \hline
Radio transceiver & CC1000\\ \hline \hline
\textbf{Energy harvester} & \textbf{Power} \\ \hline
Solar panels & $2 \times 10~\si{\watt}$\\ \hline\hline
\textbf{Battery capacity} & 5200 \si{\milli\ampere\per\hour}
\end{tabular}}
\end{table}

The marine buoy use case is preferred instead of the body sensor node use case. Indeed, the marine buoy has more energy harvesting capabilities and battery capacities than the body sensor node. The $Q$-learning algorithm is applied with the proposed reward function (Eq. \ref{eq:rewardConf}). The value of the learning rate $\alpha$ is computed using the same equation as previously (Eq. \ref{eq:alpha}), and we set the value of the discount factor $\gamma$ to $0.8$. The agent chooses a new action every $30$ minutes to let the battery state of charge change.  

\begin{figure*}[ht]
\center
\includegraphics[width=.8\linewidth]{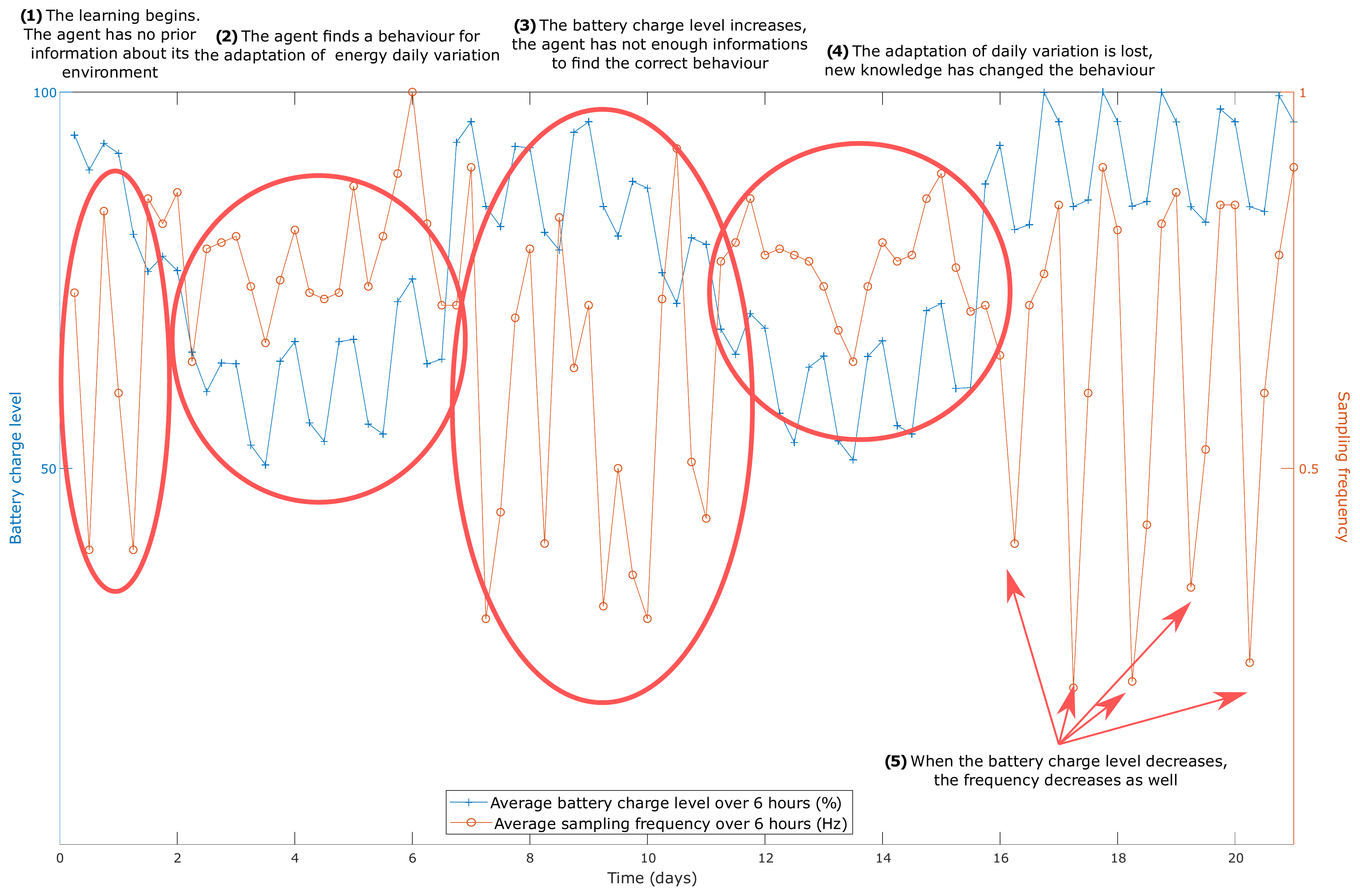}
\caption{\label{fig:percent_acq} Evolution of the battery charge level and frequency measurements of sensors using the $Q$-learning algorithm with the reward function of Equation  \ref{eq:rewardConf}. Battery capacity=$5.2~\si{\milli\ampere\per\hour}$}
\end{figure*}

Several experiments have been conducted to find the most suitable values for the parameters $\rho_{1,2,3,4}$. We found out that the best values are: $\rho_1 = 1$, ~$\rho_2 = 0.6$, ~$\rho_3 = 0.3$, ~$\rho_4 = 0$. Thus, when the battery is more than $75\%$ charged, the reward is computed only with the frequency of measurements. When the battery level is between $75\%$ and $50\%$, the reward is computed with both frequency and battery levels, but the frequency is given more weight. It is the opposite when the battery level is between $50\%$ and $25\%$, the reward function starts to preserve the battery's energy when the energy harvested is not sufficient to recharge it. If the battery charge level decreases below $25\%$, the reward is computed only with the battery level in order to preserve the node's energy. The results of a simulation using these values for the reward function (\ref{eq:rewardConf}) are presented in Figure \ref{fig:percent_acq}. 

At the beginning of the deployment (1), the agent has no prior information about its environment and takes random actions to explore it. When the battery level decreases (2), the agent adapts the frequency of measurements and finds a behaviour which adjusts the frequency according to the daily variation in the battery charge level. When the battery charge level increases around day 6 (3), the agent has not enough information in the new state to find a correct behaviour. And when the battery decreases again (4), the first knowledge learned has been replaced with the new information; the agent lost the behaviour which adjusts the frequency of measurements according to the daily variation. Nevertheless, at the end of the simulation (5) around the $16^{th}$ day, the agent's behaviour adapts correctly the frequency to the variation, this behaviour receives enough rewards to be reinforced. 

\begin{figure*}[ht]
\center
\includegraphics[width=.8\linewidth]{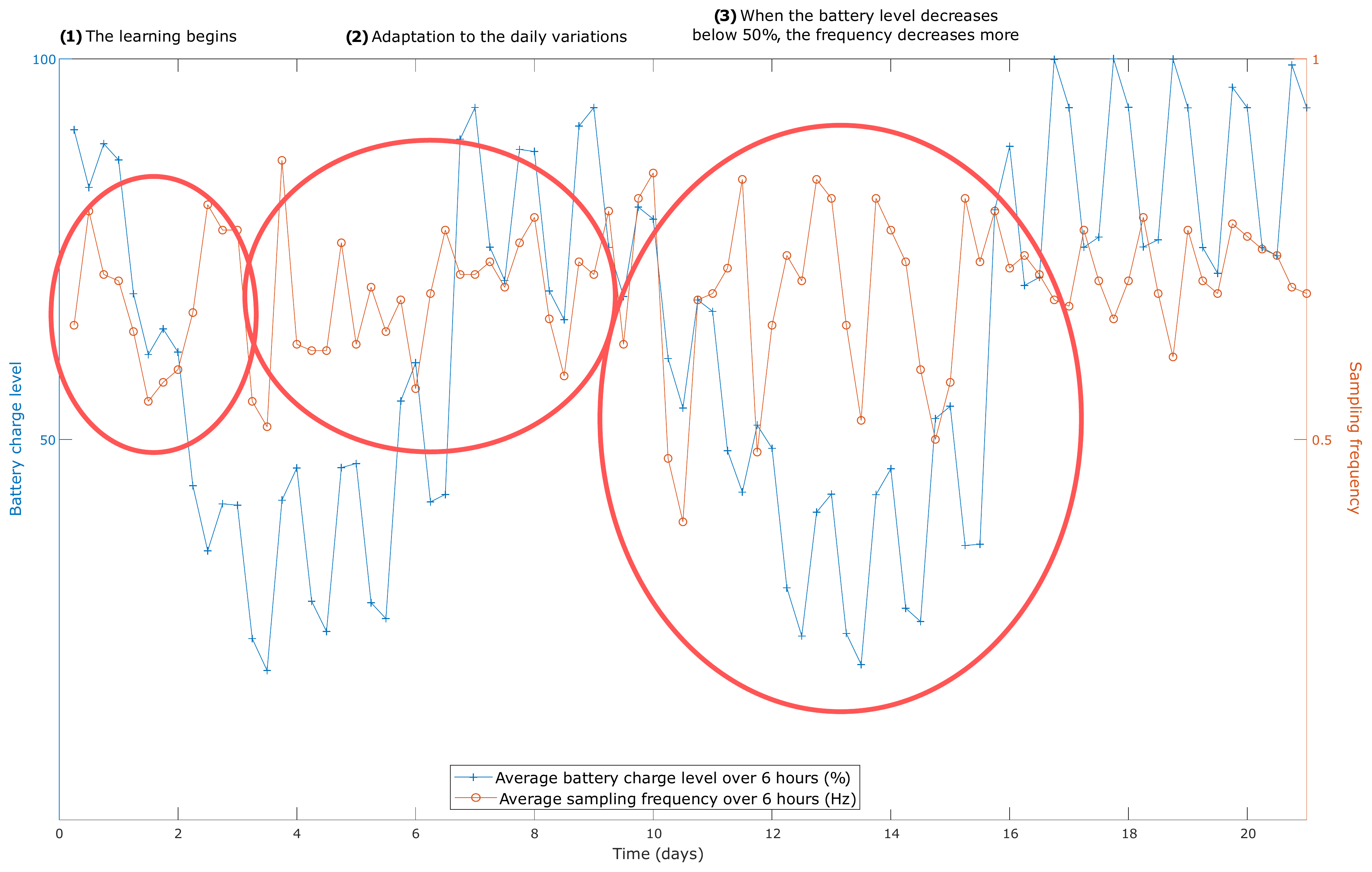}
\caption{\label{fig:percent_acq_lowBattery} Evolution of the battery charge level and measurements frequency of sensors using the $Q$-learning algorithm with the reward function \ref{eq:rewardConf}. Battery capacity=$3.2~\si{\milli\ampere\per\hour}$}
\end{figure*}

Since the battery does not decrease enough, the agent never explores the environment when the battery level is critical. So, a second experiment is conducted where the battery capacity is reduced to $3.2~\si{\milli\ampere\per\hour}$ ~instead of $5.2~\si{\milli\ampere\per\hour}$. The result of this simulation is shown in Figure \ref{fig:percent_acq_lowBattery}.

At the beginning of the simulation (1), the agent has still no prior information about its environment. Then, it successfully adapts to the daily variations in the battery charge level (2). When the battery level decreases below the $50\%$ of charge level (3), the agent decreases the frequency of measurements more than during the previous simulation. The results show that the proposed reward function is suitable for preserving the battery and for maximizing the performance when the battery is recharged. However, the agent's behaviour does not show an adaptation as good as previously. This change in behaviour is due to the exploration.  

Nevertheless, a difficulty with this reward function is to determine the values of the $\rho$ parameters. The adjustment of the parameters is based on the expertise of the designer or is determined empirically. To obtain the best behaviour with the use case, the parameter $\rho$ was determined empirically, but it can be different according to the application and the prevalence of the performance or the energy. For such an approach to be accepted, it is important to avoid adding extra-parameters to tune. In such case the reward function becomes more complex and the behaviour of the agent becomes less predictable. Thus, in the following section, we present an improvement of the proposed reward function where the parameters $\rho_{1,2,3,4}$ are removed and the different levels too. This improvement eliminates the adjustment step, making it easier to use the new reward function. Furthermore, the performance of the new reward function still remains similar to that of the previous one.

\section{\label{continuous}Continuous Reward Function to Balance the Performance and the Energy Consumption}

The selection of the different variables used in a RL approach is time-consuming and add more complexity. So, the designer needs to reduce the number of parameters to be tuned in the reward function. The previous reward function is efficient and increases the learning speed as compared to the reward function presented in the previous section (\ref{eq:rewardConf}). The values of the different parameters can be adjusted to correspond perfectly to the desired behaviour for a given application. Nevertheless, most sensor nodes are used for environmental monitoring applications, and the main objective is to extend the node's lifetime. So, in this section, we present a reward function that reinforces the same behaviour as the previous one, but without any parameter to tune.
 
Indeed, while experimenting different values for $\rho$, we observe that this parameter's value varies as the battery level. Using this observation, we design a new reward function without parameter to tune (Eq. \ref{eq:rewardConf2}) to balance the battery charge level and the performance:
\begin{equation} \tag{R7}
\label{eq:rewardConf2}
R = F_s\times B + B\times (1 - B)
\end{equation}
where $F_s$ is the frequency of the measurements and $B$ the battery level. The parameters $\rho_{1,2,3,4}$ have been replaced by the value of the battery charge level.

This reward function is a generalization of the previous one. The reward is computed mainly with the frequency of measurements when the battery is fully charged, and mainly with the battery level when the battery level is low. Thus, this proposed reward function requires no additional parameter to adjust.

We conducted a simulation on the marine buoy use case. We simulated the deployment of the buoy during a period of three weeks near Lorient, France. We applied the $Q$-learning algorithm with the same value for $\alpha$ (Eq. \ref{eq:alpha}) and $\gamma = 0.8$.

\begin{figure*}[ht]
\center
\includegraphics[width=0.8\linewidth]{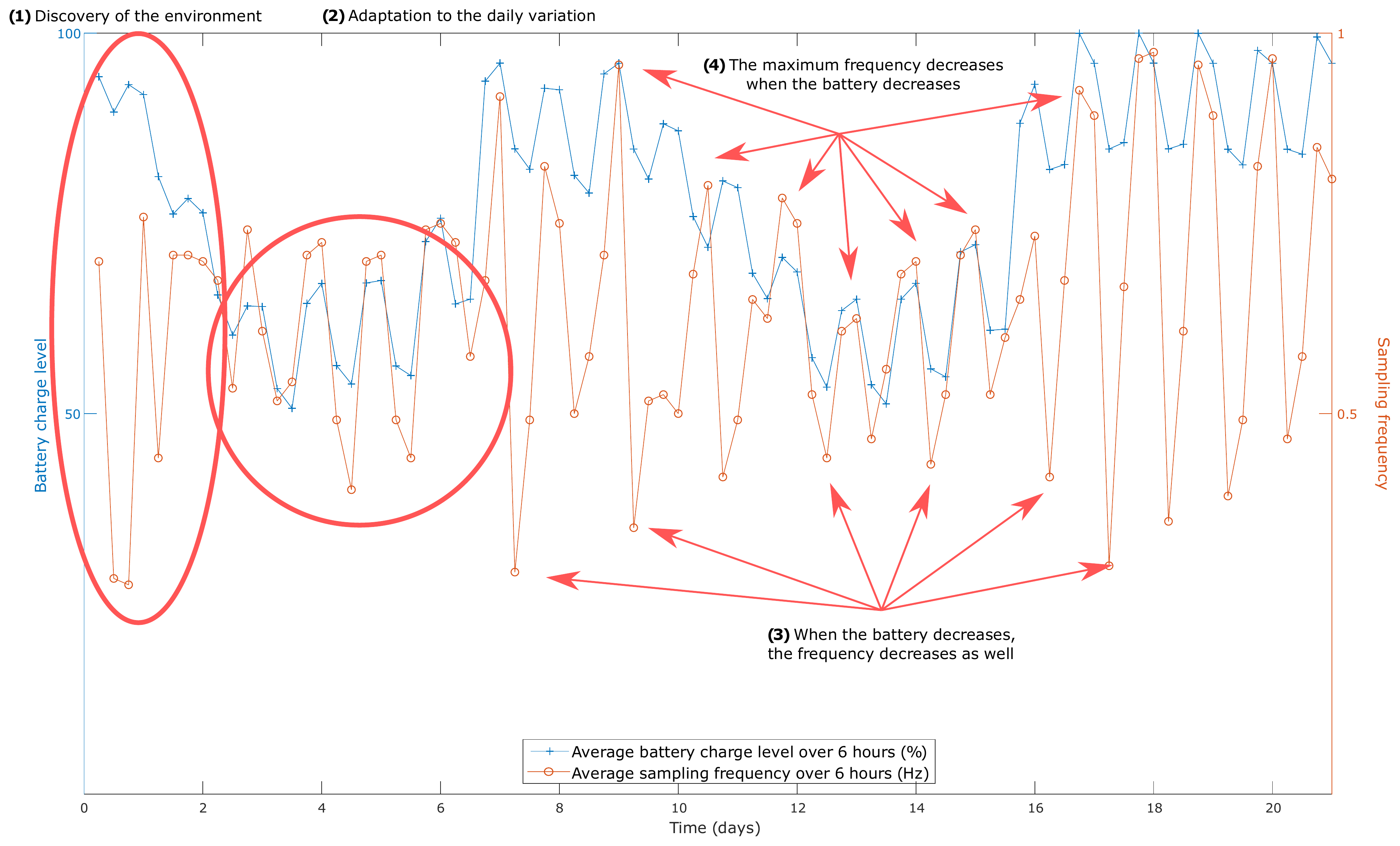}
\caption{\label{fig:loi_acq} Evolution of the battery charge level and measurements frequency of sensors using the $Q$-learning algorithm with the reward function in Eq. \ref{eq:rewardConf2}. Battery capacity=$5.2~\si{\milli\ampere\per\hour}$}
\end{figure*}

The simulation results (Fig. \ref{fig:loi_acq}) show that the agent adapts correctly the measurements' frequency to the battery's behaviour. At the beginning of the deployment (1), the battery level decreases quickly and the agent adjusts almost immediately the frequency of measurements (2). Then, when the battery level increases (4) or decreases (3) due to the harvested energy, the agent reacts and adapts the frequency of measurements accordingly. The frequency of measurements is maximum when the battery is fully charged. Before the end of the simulation, the agent achieves the correct behaviour. The agent is able to adjust the frequency of measurements to the battery charge in less than three days, when it needs more than two weeks to adapt in the previous experiments.
 
\begin{figure*}[ht]
\center
\includegraphics[width=.8\linewidth]{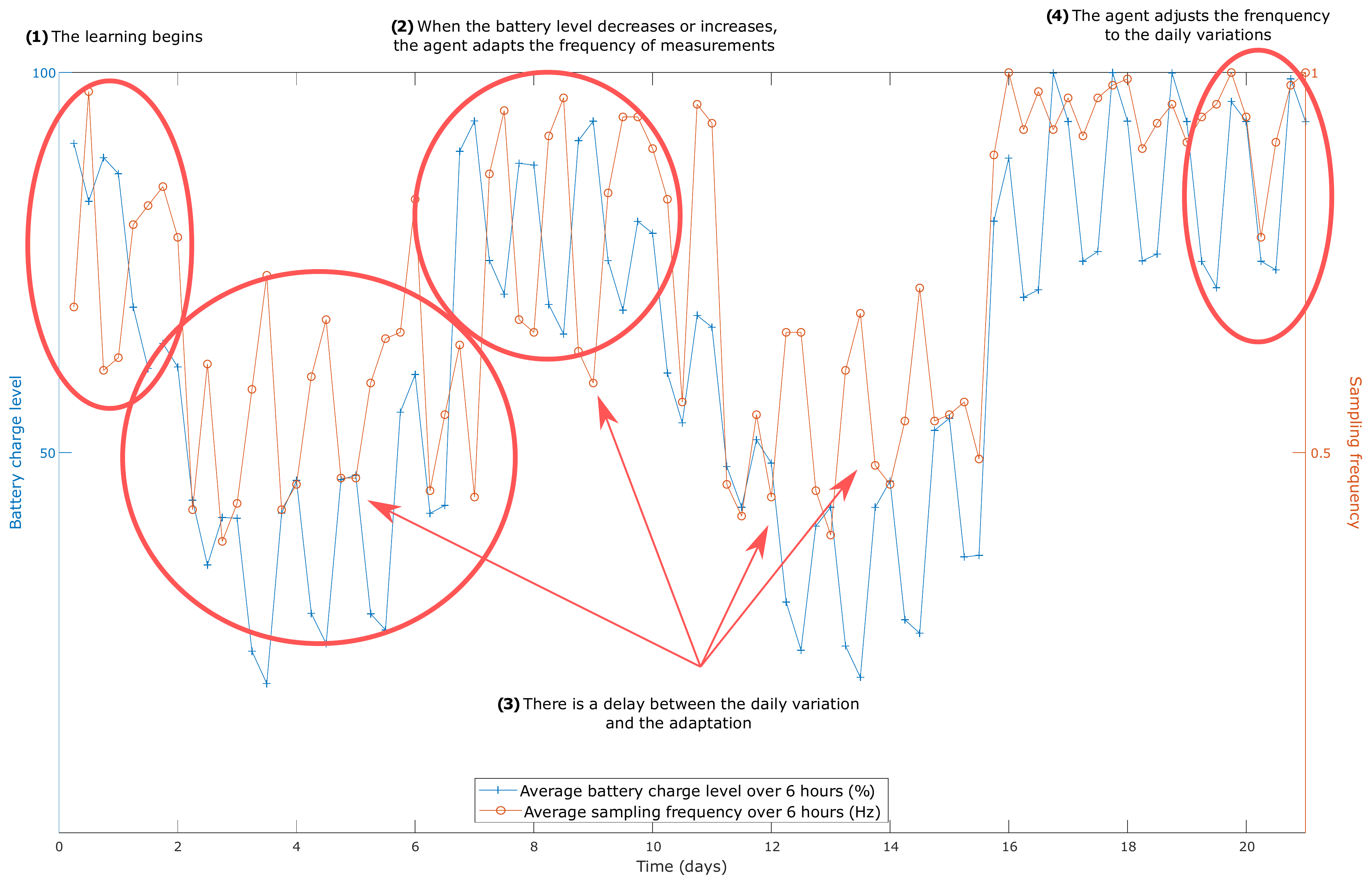}
\caption{\label{fig:loi_acq_lowBattery} Evolution of the battery charge level and measurements frequency of sensors using the $Q$-learning algorithm with the reward function in Eq. \ref{eq:rewardConf2}. Battery capacity=$3.2~\si{\milli\ampere\per\hour}$}
\end{figure*}

The proposed reward function is able to improve greatly the learning speed, the learning time is reduced by $81\%$. Furthermore, it adjusts the balance between the performance and the battery level according to the battery level without any balancing parameter. A second experiment is done with a smaller battery in the same way as for the previous reward function. The results of this experiment are shown in Figure \ref{fig:loi_acq_lowBattery}.

\begin{figure*}[ht]
\center
\includegraphics[width=.8\linewidth]{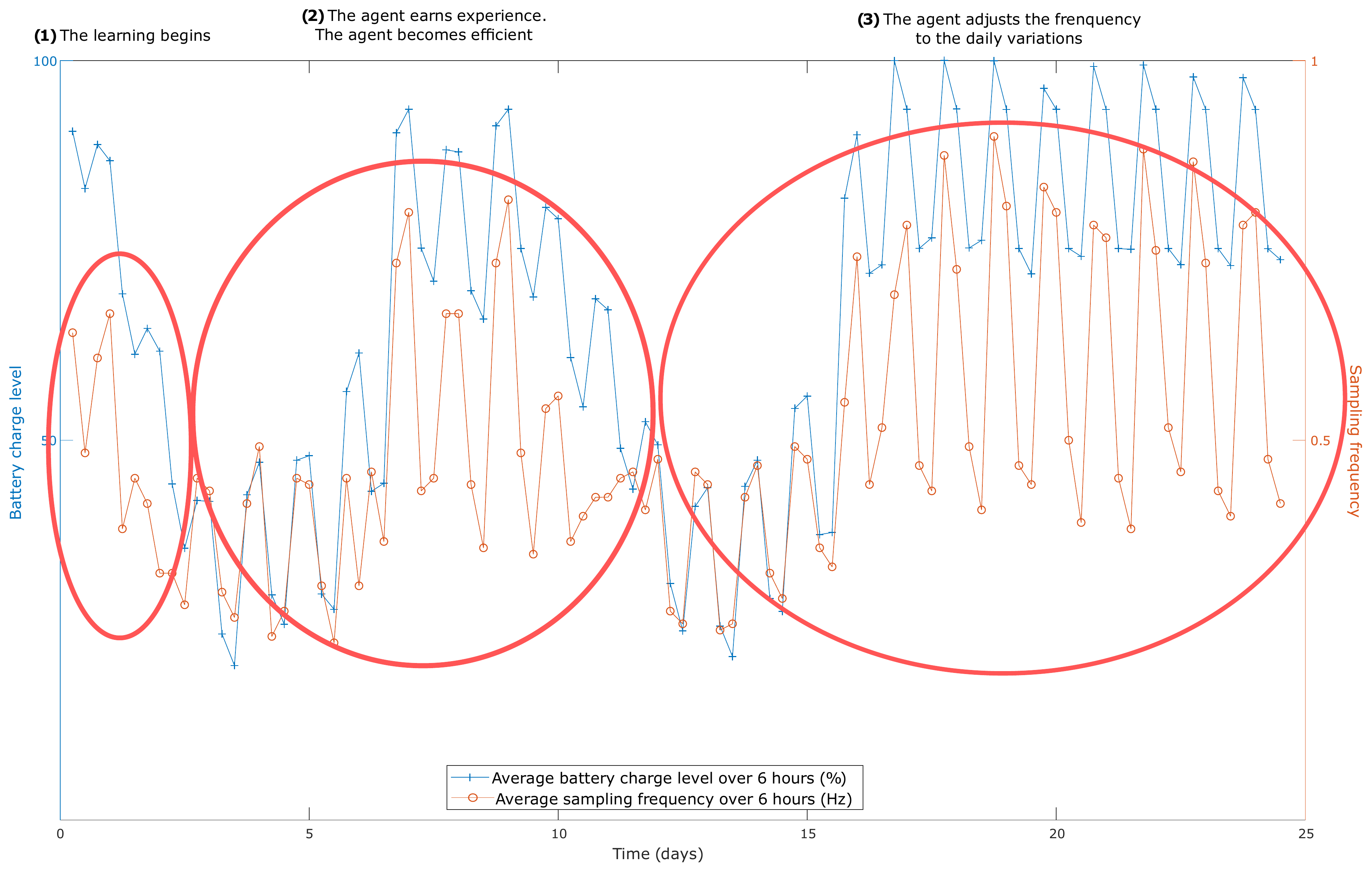}
\caption{\label{fig:loi_acq_lowBattery_24days} Evolution of the battery charge level and measurements frequency of sensors using the $Q$-learning algorithm with the reward function \ref{eq:rewardConf2}. Battery capacity=$3.2~\si{\milli\ampere\per\hour}$, but number of days=24.}
\end{figure*}

At the beginning of the deployment (1), the agent has no prior information about the environment. However, when the battery charge level decreases (2) the agent correctly adjusts the frequency of measurements, even if the daily variation are not respected (3). Then, the battery charge level increases due to the energy harvested and the agent increases the frequency as well. At the end of the simulation (4), the agent seems to respect the daily variation in the battery. To confirm the convergence of $Q$-values with the proposed reward function, we conducted a slightly longer simulation of 24 days, shown in Figure \ref{fig:loi_acq_lowBattery_24days}.

The simulation results shown in Figure \ref{fig:loi_acq_lowBattery_24days} confirms the convergence of the $Q$-values. Furthermore, the agent ends up complying with the daily variation. The differences between the two simulations are due to the exploration of the environment. During the exploration, the agent takes random actions but the agent's policy converges to the same behaviour.

When compared to the results of the previous reward function (\ref{eq:rewardConf}, Figures \ref{fig:percent_acq} and \ref{fig:percent_acq_lowBattery}), we observe that the new proposed reward function adapts more efficiently the measurements' frequency to the battery charge level. This reward function improves the performance when the battery charge level is high. Another important point is that the reward functions in Eq. \ref{eq:rewardConf} and Eq. \ref{eq:rewardConf2} are both suitable regardless of the battery capacity. The system is seen by the agent as a black box, it does not need to know the different components.

%% file: Paper/conclusion.tex
\section{\label{conclusion}Conclusions and Perspectives}
The use of energy harvesting technology creates a need for new energy management algorithms. The approach using reinforcement learning is an interesting solution for this type of problem. However, there are many obstacles to overcome such as the dilemma between exploration and exploitation, the choice of the learning rate value, and the definition of the reward function.

The reward function influences the behaviour of the node. Its design is an important part of the work and must be explained. In this paper, we experiment different reward functions in a series of simulation to identify the best design. We find out that including a balancing parameter to adjust the trade-off between performance and energy consumption is the best solution. 

These generic reward functions can achieve the same performance as a specific reward function.

We conducted a series of experiments to design a reward function able to balance energy consumption and performance. We found out that the most effective reward functions (\ref{eq:1}, \ref{eq:2}) take into account the balance with a $\beta$ parameter. They succeed in adjusting the energy consumption with the energy harvested. Another effective reward function (\ref{eq:5}) takes only into account the dominant frequency of motion. The obtained behaviour adapts to energy harvesting; however, it does not allow increasing the importance of performance on energy consumption. 

The other tested reward functions (\ref{eq:3}, \ref{eq:4}) did not allow a good energy management. Either the reward function did not make a connection between the node's consumption and the energy harvested (in particular \ref{eq:3}), or the reward function did a connection but failed to choose less consuming actions when the harvested energy is low (especially \ref{eq:4}).

The design of the reward functions \ref{eq:1} and \ref{eq:2} is interesting. We propose a modification to take account into the importance of energy consumption depending on the battery state of charge using several balancing parameters. The results show that the node adjust more efficiently its behaviour. Noticing that the value of the balancing parameters is close the battery state of charge, we propose two reward functions (\ref{eq:rewardConf} and \ref{eq:rewardConf2}) without fixed balancing parameter which perform even better than the previous ones. 

In a real application, the sensor nodes have more settings to adjust and present more parameters to balance (e.g. several sensors with different importance, several concurrent applications running on the same node), which might require adaptation of the reward functions. The Q-learning algorithm is not efficient for large Markov Decision Process. Indeed, it uses a look up table and needs to explore several times each state-action pair. Nevertheless, its neural version (Deep Q-learning) is able to scale for large environment. Future work includes reward functions taking into account a maximum of different parameters to adjust energy consumption according to the harvested energy and performance requirements. We also plan to use Deep Q-learning to cope with larger Markov Decision Processes.